\documentstyle[12pt]{article}
\begin{document}
\textheight=20 true cm
\textwidth=15 true cm
\normalbaselineskip=24 true pt
\normalbaselines
\bibliographystyle{unsrt}

\def\be {\begin{equation}}
\def\ee {\end{equation}}
\def\bea {\begin{eqnarray}}
\def\eea {\end{eqnarray}}

\def\bl{\begin{flushleft}}
\def\el{\end{flushleft}}
\def\br{\begin{flushright}}
\def\er{\end{flushright}}
\def\bc{\begin{center}}
\def\ec{\end{center}}

\def\g {\gamma}
\def\r {\rho}
\def\p {\pi}
\def\P {\Pi}
\def\G {\Gamma}
\def\a {\alpha}
\def\as {\alpha_s}
\def\az {{a^2\over 1-\zeta^2}}

\setcounter{page}{0}
\thispagestyle{empty}

\br
{\sf SINP-TNP/95-07}\\ {April 1995}
\er

\vskip 1 true cm

\bc
{\Large\bf Electroweak Precision Data\\
and a Heavy $Z'$}\\[5mm]
{\large Anirban Kundu \footnote{Present address: Theoretical
Physics Group, Tata Institute of Fundamental Research, Bombay - 400 005}
}\\[2mm]
Theory Group, Saha Institute of Nuclear Physics,\\
1/AF Bidhannagar, Calcutta - 700064, India.\\
Electronic Address: {\em akundu@theory.tifr.res.in}
\ec

\vskip 1 true cm

\begin{abstract}

We consider the physics of an extra $U(1)$ gauge boson $Z'$, 
which can mix with $Z$
through intermediate fermion loops. The loop contribution due to the
heavy top quark significantly affects the low-energy observables, and
for $m_{Z'}>m_Z$, one can always adjust the
shifts in these observables to be in the
right direction suggested by experiments, when we impose the anomaly
cancellation conditions for $Z'$.

\end{abstract}
\newpage

With the ever-increasing precision of the electroweak experiments,
some disturbing signatures about the validity of the Standard Model
(SM) are coming into view. Most notable among them are (i) $R_b =
\Gamma(Z\rightarrow b\bar b)/\Gamma(Z\rightarrow hadrons)$, (ii) the
left-right asymmetry $A_{LR}$ measured at SLAC, and (iii) the
$\tau$-polarization asymmetry, ${\cal P}_{\tau}$. At the same time,
observables such as the total $Z$-width, $\Gamma_Z$, and the hadronic
cross section at the $Z$-peak, $\sigma_{had}$, are so well measured
that arbitrary extensions of the SM are severely constrained. Among
the non-supersymmetric extensions, technicolor is struggling to make
itself compatible with the oblique electroweak parameters, $R_b$, and
the FCNC data, and is not yet convincingly successful; extra fermion
generations do not seem to resolve the discrepancies in the measured
values of the abovementioned quantities, and are also restricted by
the oblique parameters $S$ and $T$. It has been shown \cite{km} that
addition of any number of arbitrary scalar representations,
satisfying the constraints on $\rho$ and on asymptotic unitarity,
invariably worsens the discrepancy in $R_b$, and is totally
insensitive to $A_{LR}$. 

The only physically interesting
choice that remains is the addition of one or more extra
gauge bosons. Holdom \cite{holdom} and Caravaglios and Ross
\cite{ross} have already discussed that possibility in the
literature. Both of these references add an extra neutral gauge
boson $Z'$ to the SM particle spectra. While Holdom has considered a
tree-level mixing between $Z$ and $Z'$, Caravaglios and Ross have
focussed on the Born graph of $e^+e^-\rightarrow f\bar f$ mediated by
$Z'$. However, the $Z'f\bar f$ couplings derived from the experimentally
measured parameters are not free from anomaly, and thus one has to add
extra fermions to the model. These fermions not only contribute to
the oblique parameters, but may also introduce significant loop
corrections to the observables, thus making the whole pattern of the new
couplings somewhat confusing, and at the worst case, untraceable. The
oblique parameters are also affected by a tree-level $Z-Z'$ mixing. 

The important point stressed by Caravaglios and Ross is that one
needs an imaginary amplitude coming from new physics effects to give
a nonzero interference with the SM amplitude. In other words, the
real part of the new physics amplitude does not contribute to
physical observables if $|{\cal M}_{new}|^2\ll |{\cal M}_{SM}|^2$. To
satisfy this property, the authors in ref. \cite{ross} have
considered a $Z'$ nearly degenerate with $Z$ so that both $Z$ and
$Z'$ propagators are imaginary (apart from a factor of $-ig_{\mu
\nu}$). However, the $Z$ lineshape and $\Gamma_Z$, as measured at
LEP, are in such conformity with the SM that the $Z'e^+e^-$ coupling
has to be unreasonably small compared to the $Z'b\bar b$ coupling,
whose value is fixed from the measurement of $R_b$. Unless there is
some strong logic (as suggested in ref. \cite{holdom}) which forbids
$Z'$ to couple with the first two fermion generations
(in the weak eigenbasis), such a model, according to our view, seems
to be quite artificial. 

In this letter we consider what we think to be a much more realistic
scenario. We assume that there is only one neutral $U(1)$ gauge boson
$Z'$. There exists a number of models which predict such a $Z'$, though
their properties vary with the models chosen. We want to make an 
analysis which is sufficiently model-independent, except the existence 
of a $Z'$, which is the common factor among these variety of models. 
As we do not confine ourselves within a particular model, our results 
are more qualitative than quantitative and to be taken as trends. However,
in nearly all the cases, the trends are in conformity with the experimental
data.

Even in performing a general analysis, one requires some sort of a 
guideline, and fortunately, the $Z'$-physics is so well-studied  that
we have quite a few of them. For example, Langacker and Luo \cite{langacker}
have shown that a $Z-Z'$ mixing at tree-level, if exists, is bound to 
be very small (less than 1\%). Thus one does not make any great error
in neglecting the tree-level $Z-Z'$ mixing altogether; moreover, it keeps
the oblique parameters unaffected by $Z'$. Another guideline is the condition 
that $Z'$-current is to be anomaly-free, and if one does not want to
extend the fermion spectrum, it imposes some restriction on the $Z'f\bar
f$ couplings. Thus, our study will be a general one except the imposition 
of these two constraints. There also exists a mass bound on $Z'$:
for a $Z'$ with SM
couplings to the fermions, the mass limit (at 95\% CL) is 412 GeV
(from direct search in $p\bar p$ colliders) and 779 GeV (from
electroweak fit to the LEP data) \cite{pdg}. If the $Z'f\bar f$
couplings do not mimic the SM ones, these limits may not be valid
({\em e.g.}, $Z'$ which couples only to the third generation
fermions). However, there is no reason for $Z'$ to be nearly
degenerate with  $Z$, and we will drop this assumption made in
ref. \cite{ross}.

One notes that if $m_{Z'}\not= m_Z$, the only way to have a
non-vanishing interference term is to consider a $Z-Z'$ mixing
mediated by fermion loops, as shown in fig. 1. This is similar to the
well-studied $\g -Z$ mixing; while the latter effects are subtracted
from experimental measurements, the former effects are not, and so
the concerned amplitude is a coherent sum of two amplitudes: pure SM
electroweak, and that arising from new physics. As the loop
contribution is proportional to $m_f^2$, only the top loop is
considered. Note that the two-loop $Z-Z'-Z$ amplitude is real and
hence does not affect the interference term.

First, let us consider a toy model 
in which $Z'$ couples
only to the third generation. This will help us to understand the
trend. The SM amplitude of $e^+e^-\rightarrow
f\bar f$ is
\be
{\cal M}_Z=ir_1\big[\bar e(p_1)\g ^{\mu}(g^e_V-g^e_A\g _5)e(p_2)
\big] \big[\bar f(p_3)\g_{\mu}(g^f_V-g^f_A\g_5)f(p_4)\big]
\ee
and the new physics amplitude is
\be
{\cal M}_{new}=ir_2\big[\bar e(p_1)\g ^{\mu}(g^e_V-g^e_A\g _5)e(p_2)
\big] \big[\bar q(p_3)\g_{\mu}({g^q_V}'-{g^q_A}'\g_5)q(p_4)\big]
\ee
where the conventional $Zf\bar f$ vector and axialvector couplings
are denoted by $g^f_V$ and $g^f_A$ respectively, and analogous
quantities for the $Z'q\bar q$ vertex (we will always use $q$ to
denote a third generation fermion) are denoted by ${g^q_V}'$ and 
${g^q_A}'$ (thus, the $Z'q\bar q$ vertex factor is given by
$(g/2\cos\theta_W)\gamma^{\mu}({g^q_V}'-{g^q_A}'\gamma_5$). 
We neglect the QED terms in the amplitudes. 
At the $Z$-peak, one has
\be
r_1={\sqrt{2}Gm_Z\over \Gamma_Z},
\ee
\be
r_2={2G^2m_Z\over (1-\zeta^2)\Gamma_Z}f,
\ee
where $\zeta=m_{Z'}/m_Z$ (as we are not on the $Z'$-peak,
$\Gamma_{Z'}$ can be neglected), and $f$ is the two-point loop
integral given in Appendix 1. With $m_t=175$ GeV \cite{cdf} and
taking the QCD corrections into account, we get
\be
f=2.90(0.018 {g^t_V}'-{g^t_A}')\times 10^3.
\ee
With ${g^t_V}'$, ${g^t_A}'$ and $\zeta$ of the order of unity,
$|r_2/r_1|$ is of the order of 0.1, so it is justifiable to neglect
the $|r_2|^2$ contributions. We have also neglected the QCD 
and the electroweak
corrections to the internal top loop, as well as the threshold effects
of ${\cal O}(\alpha\alpha_s^2m_t^2)$, and have only taken the
corrections to the external fermions into account. This introduces
an error of at most two to three per cent and as we mainly concentrate
on the qualitative features, the approximation is a good one.
Anyway, the 
quantitative results are hardly affected.
We note that it is the massive top quark
that makes the interference amplitude non-negligible. 

The cross-section with initially polarized electron beam comes out to
be
\be
\sigma_L(\theta )=Ar_1^2(g^e_L)^2\big[(1+\cos\theta)^2T_1+(1-\cos
\theta)^2T_2\big], 
\ee
\be
\sigma_R(\theta )=Ar_1^2(g^e_R)^2\big[(1+\cos\theta)^2T_2+(1-\cos
\theta)^2T_1\big], 
\ee
where $A$ is a numerical constant ($=m_Z^2/64\pi^2$), and $T_1$,
$T_2$ are given by
\be
T_1=N_c\big[r_1(g^f_L)^2+2r_2(g^f_L{g^q_L}')\big],
\ee
\be
T_2=N_c\big[r_1(g^f_R)^2+2r_2(g^f_R{g^q_R}')\big].
\ee
In the above formulae, $N_c$ is the relevant color factor, which is 1
for leptons and $3(1+\as (m_Z^2)\pi^{-1}+1.409\as^2(m_Z^2)\pi^{-2}
-12.77\as ^3(m_Z^2)\pi^{-3})$ for quarks. The right- and the left-handed
fermion couplings are related to the vector and axialvector couplings
in the conventional way:
\be
g_V={1\over 2}(g_L+g_R),\ \ g_A={1\over 2}(g_L-g_R).
\ee

From eqs. (6) and (7), it is clear that only those observables which
involve third generation fermions in the final state will be
modified. Thus, the forward-backward electron asymmetry $A^e_{FB}$ or
the partial width $\Gamma (Z\rightarrow e^+e^-)$ retain their SM
values, while observables like $\Gamma_Z$, $A^b_{FB}$, ${\cal
P}_{\tau}$, $R_b$ (and other partial widths) will have contributions
coming from the $Z-Z'$ mixing. Low-energy observables are not
sensitive to this mixing as the $Z$-propagator, apart from
$-ig_{\mu\nu}$, is real, and the interference term vanishes. Lepton
universality is also not respected in this model. The expressions for
the modified observables follow immediately from eqs. (1) and (2);
however, they do not throw much light on the nature of the
modification, as one has to take account of seven arbitrary $Z'q\bar
q$ couplings (three in the lepton sector and four in the quark
sector). Here we impose the condition that the $Z'$ current has to be
anomaly free. This assures that no new fermions are required in the
model and eq. (5) remains unchanged. A simple way to do that is
to take the new couplings proportional to the hypercharge $Y$ of the
corresponding fermions (this is, by no means, the only choice). 
Denoting this proportionality constant by $a$, we obtain
\be
({g^{\nu_\tau}_L}',{g^{\tau}_L}',{g^{\tau}_R}',
{g^t_L}', {g^t_R}',{g^b_L}',{g^b_R}')=(-a,-a,-2a,a/3,4a/3,a/3, -2a/3).
\ee
The total $e^+e^-$ annihilation cross-section at $s=m_Z^2$ changes by
an amount $\delta\sigma$, which is also a measure of the change in
$\Gamma_Z$. With the couplings given in eq. (11), this change comes
out to be
\be
{\delta\sigma\over\sigma}={\delta\Gamma_Z\over\Gamma_Z}=-8.76  \times
10^{-3}{a^2\over 1-\zeta^2}
\ee
where we have taken $G=1.16639\times 10^{-5}$ GeV$^{-2}$,
$m_Z=91.189$ GeV and $\Gamma_Z=\Gamma_Z^{SM}=2.497$ GeV. Note that
eq. (12) is independent of the sign of $a$; this is because $Z'q\bar
q$ couplings always come in pair, one being the internal $Z't\bar t$
coupling. It depends on the sign of $\zeta$, and for
$m_{Z'}>m_Z$, the deviation is positive. From the experimental bound
\be
{\delta\Gamma_Z\over\Gamma_Z}\leq 3\times 10^{-3},
\ee
one gets
\be
-0.34\leq\az,
\ee
which, for $a=1$, yields $m_{Z'}\geq 181$ GeV. The change in the
hadronic cross-section is
\be
{\delta\sigma_{had}\over \sigma_{had}}=-5.8\times 10^{-3}\az
\ee
which is well within the allowed limit, and can be used to find the
change in $R_b$:
\bea
R_b &=& R_b^{SM}+(1-R_b^{SM}){\delta\Gamma(Z\rightarrow b\bar b)\over
\Gamma (Z\rightarrow hadrons)}\nonumber\\
&\leq& 0.2172.
\eea
The SM value of $R_b$, $0.2156$,
is for $m_t=175$ GeV and takes the two-loop
corrections induced by the heavy top quark into account \cite{erler}.
Branching fraction for charm, $R_c$, is reduced, but not very
significantly: 
\be
{\delta R_c\over R_c}\geq -0.0020.
\ee
The change in forward-backward $b$ asymmetry is small, and negative:
\be
{\delta A_{FB}^b\over A_{FB}^b}=0.0130\az
\ee
whereas for the $\tau$-lepton, the fractional change in the 
left-right asymmetry $\delta A^{\tau}_{LR}/A^{\tau}_{LR}$
is negative, and thus more than 
resolves the discrepancy of the experimental
value with the SM prediction:
\be
{\delta A_{LR}^{\tau}\over A_{LR}^{\tau}}=-0.3637.  
\ee
We note that in all these cases, the changes are in the right
direction, and more often than not, are in the right ballpark.
However, the lepton-universality breaking ratio, $\Gamma
(Z\rightarrow \tau^+\tau^-)/(Z\rightarrow e^+e^-)$, does not allow
such a high value of $a^2/(1-\zeta^2)$:
\be
{\Gamma(Z\rightarrow \tau^+\tau^-)\over (Z\rightarrow e^+e^-)}
=1-0.0387\az \leq 1.013.
\ee
Also, the effective number of light neutrino species is enhanced,
but within the allowed limit:
\be
\delta N_{\nu}=-0.0493\az\geq +0.016.
\ee
Thus, the upper bound of $a^2/(1-\zeta^2)$ is one order of magnitude
smaller than that allowed by $\Gamma_Z$. As Holdom has pointed out
\cite{holdom}, if the $Z'\tau^+\tau^-$ coupling is dominantly
vectorial in nature, the bounds obtained from the last two equations
can be evaded.

From eqs. (6) and (7), it is evident that $A_{LR}$ does not change.
This motivates us to move to our second model, where $Z'$ couples to
all the known fermions. The condition of anomaly cancellation hints
to a coupling pattern as shown in eq. (11), but the $a$'s may be
different for different generations. 
Thus, we are introducing three new parameters in this case compared to
one in the earlier case. Evidently, it will be easier to match the
experimental data by adjusting these parameters; on the other hand,
predictive power of the model will be somewhat lost.
However, there are certain model-independent facts which one should
take into account.

First, the Born graph, $e^+e^-\rightarrow f\bar f$ mediated by $Z'$,
will not contribute to the interference, and therefore the new
physics contribution to the tree-level amplitude will be suppressed
by a factor of $1/\zeta^2$. 
Second, if all the $a_i$'s ($i=1,2,3$) are same,
there will be no lepton non-universality, and it is possible to tune
the $a_i$'s in such a way that the non-universality remains within
the allowed limit, while keeping other predictions more or less
intact. Third, even for $\zeta > 1$, the shift in the total cross-section
at the $Z$-peak, $\delta\sigma_{tot}$, can be either positive or
negative.

Eqs. (6) and (7) are now modified to
\bea
\sigma_L(\theta )&=&Ar_1\big[(1+\cos\theta)^2\{(g^e_L)^2T_1+(g^e_L
{g^e_L}') T_2\}\nonumber\\
&{ }&\ \  +(1-\cos\theta)^2\{(g^e_L)^2T_3+(g^e_L{g^e_L}')T_4\}\big], 
\eea
\bea
\sigma_R(\theta )&=&Ar_1\big[(1-\cos\theta)^2\{(g^e_R)^2T_1+(g^e_R
{g^e_R}') T_2\}\nonumber\\
&{ }&\ \  +(1+\cos\theta)^2\{(g^e_R)^2T_3+(g^e_R{g^e_R}')T_4\}\big], 
\eea
where
\bea
T_1&=&N_cr_1(g^f_L)^2+2r_2(g^f_L{g^f_L}'),\\
T_2&=&2N_cr_2(g^f_L)^2,\\
T_3&=&N_cr_1(g^f_R)^2+2r_2(g^f_R{g^f_R}'),\\
T_4&=&2N_cr_2(g^f_R)^2.
\eea
First let us assume, for simplicity, $a_1=a_2=a_3=a$.
The limiting value of $a^2/(1-\zeta^2)$, as obtained from $\delta
\Gamma_Z/\Gamma_Z$, is more constrained compared to model 1:
\be
\az\geq -0.069
\ee
leading to $m_{Z'}\geq 446$ GeV for $a=1$.
Unfortunately, $\delta A_{LR}$ is negative ($=-0.0065$), and so this choice
fails to be the desired one. However, if one puts $-a_1=a_2=a_3=a$, the
total cross-section decreases (for $\zeta > 1$), and from the experimental
bound, one obtains
\be    
\delta A_{LR}=0.015 
\ee
which explains the trend of the SLAC result perfectly.

One must comment about the other observables, none of which are much 
affected, due to the highly constrained value of $a^2/1-\zeta^2$. 
The change in $R_b$, for the latter choice of $a$'s, 
is positive, and the result is in agreement with the experimental data.

Thus, both these models allow FCNC processes, forbidden in the SM. For the
second model, one needs different $a_i$'s (and thus the maximum
splitting between the $a_i$'s can be restricted). The processes now
allowed include GIM-violating $Z$-decays, and tree-level $B_d-\bar
B_d$ (and $B_s-\bar B_s$) mixing. However, 
the inherent uncertainties limit the usefulness of such processes 
in detecting a new gauge boson indirectly. 

\bigskip

In this letter, we show that the trend of
some of the present experimental data,
which may indicate a deviation from the SM, can be explained by
considering a heavy neutral gauge boson $Z'$. A crucial role is
played by the heavy top quark which ensures a significant
contribution from the interference term in the $e^+e^-\rightarrow
f\bar f$ amplitude. Two models are considered;
one in which $Z'$ couples only to the third generation fermions and
another in which it couples to all the three generations. The first
model allows a lower value of $m_{Z'}$. Guided by the anomaly
cancellation conditions of the new gauge boson, we find that the
shifts in the measured observables are {\em always} in the right
direction. We expect that these results
may motivate a search, direct or indirect, for $Z'$ in the future
colliders. 

\newpage
\centerline {\bf Appendix 1}

\bigskip

The two-point function (fig. 2), $i\Pi_{\mu\nu}$, can be written as
$$
i\Pi_{\mu\nu}(m_1,m_2,\lambda,\lambda ')={i\over 4\pi^2}\int_0^1 ~dx
\big[\Delta+\ln (\mu^2/M^2)\big]
$$
$$\times\big[2(1+\lambda\lambda')x(1-x)q_{\mu}q_{\nu}+(1+\lambda\lambda')
(-2x(1-x)q^2+m_1^2x+m_2^2(1-x))g_{\mu\nu}
$$
$$-(1-\lambda\lambda')m_1m_2g_{\mu\nu}\big], \eqno(A.1)
$$
where
$$
\Delta={1\over\epsilon}-\gamma+\ln 4\pi, \eqno(A.2)
$$
and
$$
M^2=-q^2x(1-x)+m_1^2x+m_2^2(1-x).\eqno(A.3)
$$
The vertex factors are $\gamma_{\mu}(1-\lambda\gamma_5)$ and
$\gamma_{\nu}(1-\lambda'\gamma_5)$ respectively.

Neglecting $q_{\mu}q_{\nu}$ terms (they vanish if external fermions
are massless), and putting $m_1=m_2=m$, we get
$$
f(m,\lambda,\lambda')=-{1\over 2\pi^2}\Bigg[\big(1+
\lambda \lambda'\big)\Big\{(\Delta+\ln\mu^2)\big({1\over 6}q^2-
{1\over 2}m^2\big)+q^2(I_1-I_2)+{1\over 2}m^2I_3\Big\}
$$
$$
~~~~~~~+\big(1-\lambda\lambda'\big)\Big\{(\Delta+\ln\mu^2){m^2\over
2}-{1\over 2}m^2I_3\Big\}\Bigg]\eqno(A.4)
$$
where
$$
I_1,I_2,I_3=\int_0^1~dx(x^2,x,1)\ln~M^2,\eqno(A.5)
$$
and
$$
\Pi_{\mu\nu}=ifg_{\mu\nu}. \eqno(A.6)
$$
For $m\geq q/2$, the expressions for the $I$'s are
$$
I_1={\ln m^2\over 3}-{2\over 3}\Big[{13\over 12}-{m^2\over q^2}-\Big(
{5\over 4}{m^2\over q^2}-{m^4\over q^4}-{1\over 4}\Big){2\over\eta}
\tan ^{-1}{1\over 2\eta}\Big],\eqno(A.7)
$$
$$
I_2={\ln m^2\over 2}-\Big[1-2\eta\tan ^{-1}{1\over 2\eta}\Big],\eqno(A.8)
$$
$$
I_3=\ln m^2-2+4\eta\tan^{-1}{1\over 2\eta},\eqno(A.9)
$$
where
$$
\eta=(m^2/q^2-1/4)^{1/2}.\eqno(A.10)
$$
In the text, we use the $\overline{MS}$ scheme and take the subtraction 
point $\mu=m_Z$ to obtain the numerical values.

\newpage

\centerline{\bf Figure Captions}
\bigskip

\begin{enumerate}

\item $Z-Z'$ mixing mediated by $t$ loop.

\item The two-point gauge boson vacuum polarization diagram.
\end{enumerate}
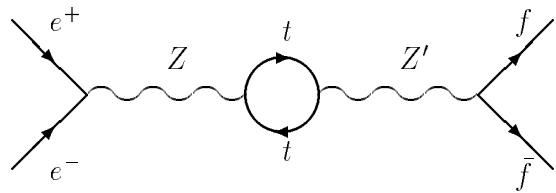
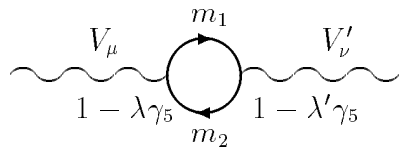
\end{document}